\documentstyle[12pt,aaspp4]{article}
\tightenlines

\def	\beq	{\begin{equation}}
\def	\eeq	{\end{equation}}
\def    \ba     {\begin{eqnarray}}
\def    \ea     {\end{eqnarray}}

\def	\H	{{\rm H}}
\def	\HH	{{\rm H}_2}
\def	\He	{{\rm He}}

\def	\d	{{\rm d}}

\def	\ba	{{\bf a}}

\def    \simlt  {\lower.5ex\hbox{$\; \buildrel < \over \sim \;$}}
\def    \simgt  {\lower.5ex\hbox{$\; \buildrel > \over \sim \;$}}

\def\pmb#1{\setbox0=\hbox{#1}%
\kern-.025em\copy0\kern-\wd0	
\kern-.05em\copy0\kern-\wd0
\kern-.025em\raise.0433em\box0}

\def \bomega {{\pmb{$\omega$}}}
\lefthead{Lazarian \& Draine}
\righthead{Disorientation of grains}

\begin{document}

\title{Disorientation of Suprathermally Rotating Grains and the 
Grain  Alignment Problem}

\author{A. Lazarian   \&  B.T. Draine}
\affil{Princeton University Observatory, Peyton Hall, Princeton,
NJ 08544}

\begin{abstract}
We discuss the dynamics of dust grains subjected to uncompensated torques
arising from H$_2$ formation. In particular, we discuss grain dynamics
when a grain spins down and goes through a ``crossover''. As first
pointed out by Spitzer \& McGlynn (1979), the grain angular momentum before
and after a crossover event are correlated, and the degree of this 
correlation affects the alignment of dust grains by paramagnetic dissipation.
We calculate the correlation including the important effects of 
thermal fluctuations
within the grain material. These fluctuations limit the degree to which
the grain angular momentum ${\bf J}$ is coupled with the grain principal axis 
${\bf a}_1$ of maximal inertia. We show that this imperfect coupling of 
${\bf a}_1$ with ${\bf J}$  plays a critical role during crossovers 
and  can substantially increase the efficiency of
paramagnetic alignment for grains larger than $10^{-5}$~cm. As a result,
we show that for reasonable choices of parameters, the observed
alignment of $a\gtrsim 10^{-5}$ cm grains could be produced by
paramagnetic dissipation in suprathermally rotating grains,
if radiative torques due to starlight were not present.
We also show that the efficiency of
mechanical alignment in the limit of long alignment times is not altered 
by the thermal fluctuations in the grain material.

\end{abstract}


\section{Introduction}

Understanding of the observed
 alignment of interstellar grains
is a  challenging problem of nearly a half century's 
standing (see Roberge 1996). Lacking 
 a proper understanding of the alignment processes, we can only tentatively
 interpret polarimetric data in terms of
the magnetic field. Indeed, polarizing grains can be aligned 
with long axes either perpendicular or parallel to the magnetic field, 
depending on what causes the alignment (see Lazarian 1994); they
can also be not aligned at all (see Goodman 1996).

One of the essential features of grain dynamics in the diffuse interstellar
medium (henceforth ISM) is suprathermal rotation (Purcell 1975, 
1979). Originally, three separate
causes of suprathermal rotation were suggested: inelastic scattering of 
impinging atoms when the gas and grain temperature differ, photoelectric
emission, and H$_2$ formation on grain surfaces. The latter was shown
to dominate the other two for typical conditions in the diffuse
ISM (Purcell 1979). More recently,
 radiative torques due to starlight have been identified as a major 
mechanism driving suprathermal rotation (Draine \& Weingartner 1996, 1997). 

Alignment of grains rotating suprathermally  differs considerably
from the
alignment of thermally rotating grains.\footnote{See 
Davis \& Greenstein (1951),
Jones \& Spitzer (1967), Mathis (1986), 
Roberge et al. (1993), Lazarian (1995a) for 
paramagnetic alignment of thermally rotating grains; and Gold (1951),
Lazarian (1994), Roberge et al. (1995), Lazarian (1996a) for
mechanical alignment of thermally rotating grains.} The theory of
paramagnetic alignment of suprathermal  grains was discussed 
by Purcell (1979) and Spitzer \& McGlynn (1979); henceforth SM), and has
been elaborated by Lazarian (1995b,c, 1996b). Until recently, 
mechanical alignment, e.g. alignment caused by
a gaseous flow,   was thought 
not to be applicable to suprathermally rotating grains, as rapid rotation
makes the grains not susceptible to such a process. However, two new
mechanisms of mechanical alignment of suprathermally rotating grains, 
namely, the ``crossover''
and ``cross-section'' mechanisms, have been suggested recently 
(Lazarian 1995d) and shown to be
effective in interstellar regions with gas-grain streaming 
(Lazarian \& Efroimsky 1996; Lazarian et al. 1996). 

The crossover event is the 
most important period in the dynamics of suprathermally rotating
grains. The H$_2$ formation sites on a grain 
surface have a finite ``lifetime'' $t_L$, which may be determined by the 
``resurfacing'' of the grain by accreted atoms (SM, Purcell 1979) or 
poisoning of active sites by oxygen (Lazarian 1995c). Because of the 
changes in the resulting torque, the spin-up has a finite
duration and this limits the paramagnetic alignment attainable.
In the case of mechanical alignment, it is during the crossovers that the 
grain is susceptible to alignment due to gaseous bombardment.  

The pioneering study of SM showed that the direction of angular 
momentum before and after crossover are correlated. However,  
SM found that the correlation was insufficient 
for the paramagnetic mechanism to achieve significant alignment within
the model they adopted.

Recent progress in the understanding of certain subtle issues of grain
dynamics has led us to reexamine the 
crossover process. SM assumed  that during suprathermal rotation 
the grain angular momentum is perfectly aligned  with the axis of
major inertia\footnote{For brevity, we refer to the 
principal axis of largest moment of inertia as the ``axis of major 
inertia''.}. This coupling arises from 
internal dissipation\footnote{The most important internal dissipation
 process is
 Barnett relaxation (Purcell 1979).} which, as known
from theoretical mechanics,  causes a spinning solid
body to rotate about its axis of major inertia, this being the state of
minimum rotational kinetic energy for fixed angular momentum. 
The assumption of perfect relaxation seems natural, as the time-scale 
for internal relaxation for suprathermally rotating grains is many orders
of magnitude less than the time-scale of the spin-up, but it is not
exact: it disregards thermal fluctuations within the grain body.

In fact, it was shown in Lazarian (1994) that due to thermal 
fluctuations, the coupling
above is never perfect (the formal theory of this
phenomenon is elaborated in Lazarian \& Roberge 1996). The component
of angular momentum perpendicular to the axis of the major inertia,
although tiny compared to the  grain angular momentum during
suprathermal rotation, is  very important in the course of a crossover. 
We therefore reconsider the SM theory of crossovers
in order to allow for the effects of thermal fluctuations. 

In \S 2 we pose the problem and present the necessary facts
concerning incomplete internal relaxation. In \S 3 we 
derive the disorientation factor accounting both for thermal fluctuations 
within the
grain material and for the effects of  gaseous bombardment. The latter
effect is of secondary importance
 in diffuse clouds but may be important in molecular
clouds.  The consequences of the incomplete disorientation
 on paramagnetic and
mechanical alignment are discussed in \S 4 and the conclusions
are presented in \S 5.

\section{The problem}

 A crossover is the event that occurs
between two sequential spin-ups when the component of $\bf J$ parallel
to the axis of major inertia passes through zero. This is a critical
period for grain dynamics, and during the crossover the grain
is susceptible to disorientation, which will limit the effectiveness of
paramagnetic alignment. If the grain is situated in a region with
substantial gas-grain streaming, the grain is susceptible to mechanical
alignment during crossovers.
Our task in the present paper is to describe the evolution of 
grains through  crossover events, accounting for the effects of
 thermal fluctuations
within the grain material.

To understand the crossover one needs to recall certain basic features 
of suprathermal rotation. Here we assume that 
grains are spun-up by torques arising from H$_2$ formation, 
and consider a ``brick'' with dimensions $b\times b\times a$ and density
$\rho_s$. The ratio $r\equiv b/2a$ determines the degree of grain 
oblateness;  $r=1$
for the grain $2:2:1$ discussed in Purcell (1979). 
It is possible to show (SM; Draine \& Lazarian 1996) that
the components of the torque perpendicular to the axis of major
inertia average out and therefore only the component of the torque
 parallel to this axis matters. We direct the
$z$-axis along the axis of major inertia. We let
\beq
n_{\rm H}\equiv n({\rm H})+2n({\rm H}_2)~~~,
\eeq
where $n(\H)$ and $n(\H_2)$ are
 the concentrations of atomic
and molecular hydrogen, respectively; the H$_2$ fraction we
denote $y\equiv 2n(\H_2)/n_{\H}$.

The number of H$_{2}$ molecules ejected per second from an individual
site is $\gamma a^{2}n_{\H}v_{\H}(1-y)r(r+1)\nu^{-1}$, where $\gamma$ 
is the fraction of H atoms (with mean speed $v_{\H}$ and mass $m_{\H}$)
 adsorbed by the grain, and $\nu$ is the number of active sites 
over the grain surface. 
The mean square torque from H$_{2}$ formation is (see Appendix A)
\beq
\langle L_{z}^{2}\rangle \approx
\frac{2}{3}\gamma^{2}a^{6}(1-y)^2 n_{\H}^2m_{\H}v_{\H}^{2}E\nu^{-1}r^4(r+1)~~~,
\label{3}
\eeq
where $E\approx 0.2$~eV is the kinetic energy of a nascent H$_2$ molecule. 
The fluctuating torque 
$L_z$ spins up grains to an rms angular velocity
\beq
\langle\omega^2\rangle^{1/2} = 
\langle L_{z}^{2}\rangle^{1/2}\frac{t_{d}}{I_{z}}
\left(\frac{t_{L}}{t_{L}+t_d}\right)^{1/2}~~~,
\label{4}
\eeq
(Purcell 1979), where $I_{z}=\frac{8}{3}\rho_s a^5r^4$ is the $z$ 
component of the momentum of inertia, 
$t_{d}$ is the rotational damping time (see Appendix A) 
\beq
t_{d}=\frac{2r}{(r+2)}
\frac{a\varrho_{s}}{n_{\H}m_{\H} v_{\H}}\frac{1}{(1.2-0.292y)}~~~,
\label{5}
\eeq
and $t_L$ is the lifetime of an active site.

To obtain both characteristic
numerical values and functional dependencies we will use quantities 
normalized by their standard values (see Table 1). We denote 
the normalized values by symbols with hats, e.g. 
$\hat{a}\equiv a/( 10^{-5}~{\rm cm})$, with 
$ 10^{-5}~{\rm cm}$ as the standard value of grain size. 
We consider `standard' an H$_2$ formation efficiency 
$\gamma=0.2$ and $v_{\H}= 
1.5\cdot 10^{5}$~cm~s$^{-1}$. For diffuse clouds we assume that all 
 hydrogen there is in atomic form and therefore $y=0$. 
Note that sometimes the choice of `standard' values 
is somewhat 
arbitrary, e.g. for the time being  we assume the density of active 
sites\footnote{The density of active sites depends on the interplay of
the processes of photodesorption and poisoning. It is shown in Lazarian
(1995c, 1996b) that the poisoning intensifies when the number of
active sites becomes greater than a critical number. The latter number
depends on the migration time-scale of hydrogen and the activation barrier
for the reaction between physically and chemically adsorbed hydrogen 
atoms. As the details of the grain chemistry are poorly known the critical
number of active sites is highly uncertain. Therefore, for the sake of
simplicity, we do not discuss differences in poisoning of
active sites whenever their number exceeds the critical number and assume,
following SM, a constant surface density $\alpha$ of active sites.}   
$\alpha_{\H2}$ to be $10^{11}$~cm$^{-2}$, so that 
$\nu=80\hat{\alpha}_{\H2}\hat{a}^2r(r+1)$.

Using standard values of the parameters we obtain  the 
following expression for the angular velocity
\begin{eqnarray}
\langle \omega^2\rangle^{1/2} 
&\approx&
\left(\frac{3E}{\alpha m_{\H}}\right)^{1/2}\frac{\gamma}{2a^2}
\frac{(1-y)}{(1.2-0.292y)}\frac{1}{r^{3/2}(r+2)}
\left(\frac{t_L}{t_d+t_L}\right)^{1/2}
\nonumber\\
&\approx&
5.4\cdot 10^{9}~\frac{\hat{E}\hat{\gamma}}{\hat{\alpha}^{1/2}\hat{a}^2}
\frac{(1-y)}{(1.2-0.292y)}\frac{1}{r^{3/2}(r+2)}
\left(\frac{t_L}{t_d+t_L}\right)^{1/2} \,\, {\rm s}^{-1}.
\label{5b}
\end{eqnarray}
The lifetime of an active site of H$_2$ formation is limited by
both accretion of a mono-layer of refractory material
(SM) and poisoning by atomic 
 oxygen (Lazarian 1995c). The former is usually slower than the latter and
could provide long-lived spin up with $t_L\gg t_d$. Further on we
use the term  ``resurfacing'' to refer to the fastest mechanism of the two.
The component of the mean torque along the axis of major inertia
before and after, say, resurfacing may be directed either
in the same direction  as before the process, or in the opposite direction.
In the latter case the grain undergoes a spin-down. 

The mean interval 
between crossovers is (Purcell 1979)
\beq
\bar{t}_z\approx \pi (t_L t_d)^{1/2}~~~.
\label{pur}
\eeq
The active site lifetime $t_L$ is very uncertain. 
In our numerical examples in this paper we will take $t_L=10^{12}$ s. 
For our standard parameters (Table~2) in a diffuse HI cloud this corresponds
to $t_L/t_d=0.25/\hat{a}$, and $t_z\approx 1.6 \hat{a}^{1/2}t_d$.

When a grain rotates about an arbitrary axis, the angular velocity precesses
in grain body coordinates. The Barnett effect produces a magnetic moment 
parallel to the angular velocity. Purcell was the first to realize 
that this should result
in internal dissipation with a dissipation time-scale inversely 
proportional to the angular velocity squared (Purcell 1979). It is possible
to show (see \S 3.2) that this effect
 suppresses rotation around any
axis but the axis of major inertia on a time-scale $t_B\sim
10^{7}/\aleph$~s, where $\aleph$ is the ratio of grain rotational energy 
to the equipartition energy $\sim kT$.  Both H$_2$ formation (Purcell 1979)
and radiative torques (Draine \& Weingartner 1996a) can produce $\aleph>10^3$
and therefore grains tend
to rotate around their 
axes of major inertia.

Although the ratio $t_B/\bar{t}_z\approx t_B/t_d$ can be as small as 
$10^{-5}$ (see Table~2) the alignment 
of $\bf J$ with the axis of major inertia (${\bf a}_1$) is not perfect.
The deviations of ${\bf a}_1$ from $\bf J$ arise from  thermal fluctuations
within the grain material (Lazarian 1994, Lazarian \& Roberge 1996). 
To estimate the value of such
deviations recall that rotation about ${\bf a}_1$
corresponds to the minimum of the grain kinetic energy
for fixed $J$ (internal dissipation does not alter
$J$)\footnote{If  $J$ does change due to H$_2$ torques
one should substitute its value averaged over the time of internal relaxation
in Eq.(\ref{ek1}).}. For a symmetric
oblate grain with $I_z>I_x=I_y\equiv I_{\bot}$  (i.e., our ``brick'' with
$b>a$), the grain kinetic energy is
\beq
E_k(\beta)=\frac{J^2}{2I_z}\left(1+\sin^2
\beta\left(\frac{I_z}{I_{\bot}}-1\right)\right)~~~,
\label{ek1}
\eeq
where $\beta$ is the angle between $\bf J$ and ${\bf a}_1$. 
In thermodynamic equilibrium the fluctuations
of the kinetic energy 
should have a Boltzmann distribution:
\beq
f(\beta){\rm d}\beta={\rm const}\cdot\sin\beta\exp\left(-\frac{E_k(\beta)}{kT_d}\right){\rm d}\beta~~~,
\label{r1}
\eeq
where $T_d$ is the dust temperature.
It follows from (\ref{r1}) that the fluctuating component of angular momentum
perpendicular to the axis of the major inertia $\langle J_{\bot}^2
\rangle\ll J^2$ can be approximated 
\beq
\langle J_{\bot 0}^2 \rangle \approx 
\left(\frac{I_zI_{\bot}kT_d}{I_z-I_{\bot}}\right)~~~.
\label{bot}
\eeq

We may define the ``thermal transverse angular velocity''
\beq
\omega_{\bot 0}\equiv \frac{\langle J^2_{\bot 0}\rangle^{1/2}}{I_{\bot}}=
\left( \frac{I_z k T_d}{I_{\bot}(I_{z}-I_{\bot})}\right)^{1/2}=
1.05\times 10^5 \frac{\hat{T}_d^{1/2}}{\hat{\rho}^{1/2}\hat{a}^{5/2}}
\left(\frac{15}{16 r^4-1}\right)^{1/2}~~{\rm s}^{-1}. 
\eeq
As we will see below, $\omega_{\bot 0}$ is the characteristic value for the
minimum value of the grain angular velocity during a crossover.

When the rotation is suprathermal,
$\langle J_{\bot}^2 \rangle^{1/2}$
is negligible  compared to $J$  and angle
$\beta$ is very close to zero. However, as the component of
angular momentum $J_{\|}$ parallel to the axis of major inertia 
 decreases during crossovers, the angle $\beta=
\arctan(J_{\bot}/J_{\|})$ 
increases.

When the angular velocity  decreases sufficiently, internal relaxation
becomes less efficient and the value of 
$\langle J_{\bot}^2 \rangle$ rises as
a result of the stochastic character of H$_2$ formation and impacting
gas atoms. At some point
the component of $\bf J$ parallel to $\bf z$ passes through zero and the
grain flips over (SM). Our task is to calculate the correlation of the
grain angular momentum before and after the crossover. This is done in
the next section.

\section{Crossovers}

In our treatment below we repeat the reasoning introduced
in SM but with allowance for thermal fluctuations.
The zeroth approximation, following  SM,
is the dynamics of a grain subjected to regular torques only.
 The dynamical effects of the stochastic
 torques can be evaluated by an approximate 
theory based on small perturbations of the zeroth-order solution.

\subsection{Zeroth approximation}

Let ${\bf xyz}$ be a coordinate system frozen into the grain, with 
${\bf z}$ along the grain axis of major inertia ${\bf a}_1$. 
Let ${\bf x_0y_0z_0}$ be an 
inertial coordinate system, with ${\bf z}_0\|{\bf J}$ (at some 
initial time). Let $\beta$ be the angle between the  $\bf z$-axis
 and ${\bf J}$: $J_z=J\cos\beta$. If no external torques 
act, then ${\bf J}=$~constant and the $\bf z$-axis  and the angular 
velocity $ \bomega$ will each precess around ${\bf J}$ (or $\bf z_0$) 
at a frequency $\omega_p=(I_z-I_x)J_z/I_xI_z$.

Now consider the effect of a (weak) torque $\bf L$ which is fixed 
in body coordinates $\bf xyz$. On time scales long compared to 
$\omega_p^{-1}$ the rotation of the grain around $\bomega$ and the 
precession of $\bomega$ around $\bf z_0$ imply that the only torque 
component which does not average to zero is that due to $L_z$, the 
component of $\bf L$ along the $\bf z$-axis. After this averaging 
we obtain
\begin{equation}
\frac{\d {\bf J}}{\d t}=L_z\cos\beta\frac{\bf J}{|J|}.
\label{e.9}
\end{equation}
From the Euler equations (see SM) 
we find the components of $\bomega$ in body coordinates
\begin{eqnarray}
\omega_z &=& \frac{J_z}{I_z} = \frac{L_z t}{I_z},\\
\omega_{\bot}&=&J^2_{\bot}/I^2_{\bot}={\rm const},
\label{*}
\end{eqnarray}
where $t=0$ at the moment of crossover. Eq.~(\ref{e.9}) can be rewritten
\beq
\frac{\d J_{z_0}}{\d t}=L_z\cos\beta =L_z\frac{J_z}{J_{z_0}}~. 
\label{JZ}
\eeq
Since $J_z=L_z t$ we obtain
\beq
J^2_{z_0}=L_z^2t^2+J^2_{\bot}~~~.
\label{e.12}
\eeq
According to (\ref{e.9}), the direction of $\bf J$ does not change  -- 
the torque $\bf L$ acts only to change its magnitude 
(see Eq.~(\ref{e.12})). Thus the zeroth approximation 
predicts a perfect correlation between the angular momentum directions
prior to and after the crossover. The stochastic torques make the story
more involved.

\subsection{Crossovers \& Barnett fluctuations}

Our considerations above ignored the fluctuations associated with
the Barnett effect. As discussed by Lazarian \& Roberge (1996), angle 
$\beta$ fluctuates on the time of the Barnett relaxation (Purcell 1979):
\beq
t_{B}(\omega)=\frac{A_{a}}{\omega^{2}}~~~,
\label{snew31}
\eeq
where 
\beq
A_{a}=7.1\times 10^{17}\hat{a}^{2}\hat{\varrho}\hat{T_d}
\hat{K}_0(r)~{\rm s}^{-1}~~~.
\label{new33}
\eeq
and
\beq
\hat{K}_0(r)=\frac{3}{125}\frac{(4r^2+1)^3}{r^2(4r^2-1)} ~~~.
\eeq
Note that the ratio $r=1/2$ corresponds to a cubical grain, for which no
internal relaxation is expected in agreement with Eq.~(\ref{new33}).

The fluctuations in $\beta$
 span the interval $(0,\pi)$ when $J \rightarrow 
J_{\bot 0}$ (Lazarian \& Roberge 1996).
 SM showed that during crossovers $J \sim 
J_{\bot}$ and therefore such fluctuations must be accounted
for provided that $t_c>t_B(\omega_{\bot})$, where the crossover time is 
\beq
t_c=\frac{2 J_{\bot}}{\dot{J}} ~~~,
\eeq
where $\dot{J}$ is the time derivative of $J$.

When $t_B(\omega_{\bot})\ll t_c$, Barnett fluctuations will cause $\beta$ to range over the
interval $(0,\pi)$, resulting in frequent reversals of the
torque in inertial coordinates. Consequently, the actual
time spent during the crossover will be increased. Quantitative
analysis of this regime is beyond the scope of the present paper;
it does appear clear, however, that each crossover will be accompanied
by substantial disalignment when $t_B\ll t_c$.

In our present study we confine ourselves to the other limiting case, namely,
$t_c/t_B\ll 1$,  in which case the Barnett fluctuations 
during a crossover can be
disregarded and  the initial distribution
of $J_{\bot}$ with the mean value given by Eq.~(\ref{bot})
is produced by the
Barnett fluctuations during the long time interval between crossovers.

We shall prove below that for typical interstellar conditions 
the value of $J_{\bot}$ mostly arises from thermal fluctuations
within the grain material during intervals of suprathermal rotation
and therefore is given by  Eq.~(\ref{bot}).
Thus we can estimate $t_c$:
\beq
t_c\approx \frac{2 \langle J_{\bot 0}^2 \rangle}{\dot{J}}\approx
2.6\times 10^{9}\left(\frac{\hat{\rho}\hat{T}_d \hat{a}\hat{\alpha}}
{\hat{E}}\right)^{1/2}
\frac{1}{\hat{\gamma}\hat{n}_H \hat{v}_H (1-y)}\hat{Z}_0(r)~{\rm s}~~~,
\eeq
where 
\beq
\hat{Z}_0(r)=\left(\frac{3r(1+4r^2)}{5(4r^2-1)}\right)^{1/2}~~~.
\eeq

The ratio
\beq
\frac{t_B}{t_c}=\left(\frac{a}{a_c}\right)^{13/2}~~~,
\label{tbtc}
\eeq
where $a_c$ is the critical radius $a_c$, 
which  for $\omega\approx J_{\bot}/I_z$ is equal to
\beq
a_c\approx 1.47\times 10^{-5}\left(\frac{\hat{T}\hat{\alpha}}{(1-y)^2 \hat{n}_H^2 
\hat{v}_H^2 \hat{E} \hat{\rho}_s^3}\hat{K}_1(r)\right)^{1/13}~{\rm cm}~~~,
\label{a}
\eeq
where
\beq
\hat{K}_1(r)\approx\frac{234375 r^5}{(4r^2-1)(4r^2+1)^7}~~~.
\eeq

It follows from our discussion above that we attempt to deal only
with the case $a>a_c$ while leaving the more complex regime
$a<a_c$, where Barnett fluctuations during the crossover are important,
to be dealt with elsewhere.
We use the inequality $a>a_c$ rather than $a\gg a_c$
due to the strong dependence of the time ratio $t_c/t_B$ on $a$: for
$a=10^{-5}$ cm $t_c>12 t_B$, while for $a=2.0\times 10^{-5}$ cm 
$t_B>7 t_c$. Below we analyze the implications of the critical size $a_c$ in 
the context of the variations of alignment with grain size. 

The numerical value $a_c\approx 1.5\times 10^{-5}$ cm is quite robust:
the most uncertain grain parameter is the surface density of active sites
$\alpha$, but even varying $\alpha$ by a factor $10^2$ changes
$a_c$ by only a factor $10^{2/13}\approx 1.36$.

Although so far we have considered only  suprathermal
rotation driven by H$_2$ formation, 
the existence of a critical size seems to be generic to the
problem of disorientation in the course of crossovers. 

In our treatment above we disregarded gaseous friction. This is a good
approximation in the  diffuse medium 
since $t_c \ll t_d$ (see Table~2). In molecular clouds as
$y \rightarrow 1$ the two time scales may become comparable (e.g. if the
atomic hydrogen fraction $(1-y)$ drops below $10^{-3}$),
and gas drag should be included. We also assume $t_L \gg t_c$; in the case of 
$t_L \ll t_c$
it becomes important to allow for variations in the time-averaged torque
$L_z$ during the crossover.

\subsection{Random torques}

We consider the dynamical evolution given by 
Eq.~(\ref{e.12}) as a zeroth-order solution of the problem, and the  
 dynamical effects produced by stochastic torques as perturbations of 
this solution. 

Each torque event produces an impulsive change of angular momentum 
$\bigtriangleup {\bf J}=\bigtriangleup J_z {\bf z} +
\bigtriangleup {\bf J}_{\bot} $. The angular deviations of ${\bf J}$ in the 
$({\bf J},{\bf z})$-plane are given by
\beq
\bigtriangleup \eta_{\|}=
\frac{-\bigtriangleup J_{z}\sin\beta+\bigtriangleup J_{\bot 1}\cos\beta}
     {J}
\eeq
and in the transverse direction by
\beq
\bigtriangleup \eta_{\bot}=
\frac{\bigtriangleup J_{\bot 2}}{J}~~~,
\eeq
where $\bigtriangleup J_{\bot 1}$ and $\bigtriangleup J_{\bot 2}$ are the 
components of $ \bigtriangleup {\bf J}_{\bot}$
in the plane  
parallel and perpendicular to the $({\bf J},{\bf z})$-plane, respectively. 
 The grain is subject to the action of various torques. 
Here we discuss only torques arising from H$_2$ formation 
$(\bigtriangleup J)_{\H}$ and gaseous bombardment, $(\bigtriangleup J)_g$. 
Let $N_1$ be the rate of H$_2$ formation, and $N_2$ be the rate of 
gas-grain collisions.
To simplify our notation, we denote the mean square change in angular
momentum per H$_2$ formation event
\beq
(\bigtriangleup J_i)^2=(\bigtriangleup J_i)_{\H}^2+
\frac{N_2}{N_1}(\bigtriangleup J_i)_g^2 ~~~,
\label{br17}
\eeq
where $N_2/N_1$ is the number of gas-grain collisions per  H$_2$ 
formation event. 
The mean quadratic deviation of ${\bf J}$ per H$_2$ formation is 
\beq
\langle \bigtriangleup (\eta)^{2}\rangle=
J^{-2}\{ \langle (\bigtriangleup J_{z})^{2}\rangle\sin^{2}\beta +
\langle (\bigtriangleup J_{\bot 1})^{2}\rangle\cos^{2}\beta+ 
\langle (\bigtriangleup J_{\bot 2})^{2}\rangle-
\langle \bigtriangleup J_{z}\bigtriangleup J_{\bot 1}\rangle\sin2\beta\}~~~,
\eeq
where $\langle \bigtriangleup J_{z}\bigtriangleup J_{\bot 1}\rangle=0$
due to rotation around the axis of major inertia. For a symmetric 
oblate grain 
\beq
\langle (\bigtriangleup J_{\bot 1})^{2}\rangle=
\langle (\bigtriangleup J_{\bot 2})^{2}\rangle=
\frac{1}{2}\langle (\bigtriangleup J_{\bot})^{2}\rangle~~~,
\eeq
and we obtain
\beq
\langle (\bigtriangleup \eta)^{2}\rangle=J^{-2}\left\{
\langle (\bigtriangleup J_{z})^{2}\rangle\sin^{2}\beta +
\langle (\bigtriangleup J_{\bot })^{2}\rangle\left(\frac{1+\cos^{2}\beta}{2}
\right)\right\}~~~.
\label{br20}
\eeq
The cumulative deflection due to $i+1$ random impulses may be expressed as 
\beq
\cos \eta_{i+1}=
\cos\eta_{i}\cos(\bigtriangleup\eta_{i})+
\sin\eta_{i}\sin(\bigtriangleup\eta_{i})\cos\phi_{i}~~~,
\eeq
where $\phi_{i}$ is the angle between the deviation $(\bigtriangleup\eta_{i})$ 
and the great 
circle measured by $\eta_{i}$. Averaging provides us with the result 
(see SM)
\beq
\langle\cos\eta_{f}\rangle=\Pi_{i}\langle\cos\bigtriangleup\eta_{i}\rangle~~~.
\eeq
As $\bigtriangleup\eta$ is small, we can expand
 $\cos \bigtriangleup\eta$ to obtain 
\beq
\Pi_{i}\langle \cos(\bigtriangleup\eta_{i})\rangle\approx 
\Pi_{i}\left(1-\frac{1}{2}\langle(\bigtriangleup\eta_{i})^{2}\rangle
\right)\approx 
\Pi_i\exp\left(-\frac{1}{2}\langle(\bigtriangleup\eta_{i})^{2}\rangle
\right)~~~,
\eeq
Hence
\beq
\langle\cos\eta_{f}\rangle\approx \exp(-F)~~~,
\label{m17}
\eeq
where $F$ is the disorientation parameter
\beq
F\equiv\frac{1}{2}\int_{-\infty}^{+\infty}N_{1}
\langle(\bigtriangleup\eta)^{2}\rangle{\rm d}t
\label{m18}
\eeq
and $N_{1}=L_z/\langle \bigtriangleup J_z\rangle$ 
is the number of H$_2$ torque events per second. From Eq.~(\ref{br17}) and
(\ref{br20}) it is seen that
both H$_2$ torques and gaseous bombardment are included in 
Eq.~(\ref{m18}). 

To evaluate $F$, we obtain $\beta(t)$ from the zeroth-order grain 
dynamics with only regular torques ($J_{\bot}=$const, $J_z=L_z t$);
from $\tan\beta=J_{\bot}/J_z$ we obtain
\beq
{\rm d}t=\frac{{\rm d}J_{z}}{L_{z}}=-\frac{J_{\bot}{\rm d}\beta}
                                         {L_{z}\sin^{2}\beta}=
-\frac{J^{2}}{J_{\bot}L_{z}}{\rm d}\beta ~~~.
\label{m19}
\eeq
Substituting this into Eq.~(\ref{m18}) and integrating from $\beta=0$ to 
$\beta=\pi$, one 
obtains
\beq
F=\frac{\pi}{4}\frac{\langle(\bigtriangleup J_{z})^{2}\rangle}
                    {J_{\bot}|\langle\bigtriangleup J_{z}\rangle|}
\left(1+\frac{3}{2}\frac{\langle(\bigtriangleup J_{\bot})^{2}\rangle}
			{\langle(\bigtriangleup J_{z})^{2}\rangle}\right)~~~.
\label{f27}
\eeq
Using Eq.~(\ref{br17}) and (\ref{ap6}-\ref{ap10}), we find
\beq
F=\frac{179\pi}{224}\frac{\langle(\bigtriangleup J_{z})^{2}\rangle}
                    {J_{\bot}|\langle\bigtriangleup J_{z}\rangle|}\hat{K}_1(r)~~~,
\label{f-fact}
\eeq
where
\beq
\hat{K}_2(r)=\frac{56}{179}\left[1+\frac{3[16r^2(r+1)+6r+3]}{8r^2(2r+5)}
\right]~~~.
\eeq

Eq.~(\ref{f27}) was derived assuming $J_{\bot}= $~const; in fact, it is the
value of $J_{\bot}$ when $\beta\approx \pi/2$ which should be used
in Eq.~(\ref{f27}). 
In the pioneering study by SM, it was assumed that 
$J_{\bot}$ is initially zero. This assumption is
valid only for grain temperatures  approaching absolute zero.
For nonzero grain temperatures the mean value of this component
squared cannot be less than $J_{\bot 0}^2$ given by Eq.~(\ref{bot}).
To account for this non-zero component, while avoiding solving the
corresponding Fokker-Planck equation, in our simplified treatment here
we consider the evolution of the difference $\langle J_{\bot}^{2}-
J_{\bot 0}^2\rangle$, using the lucid 
approach suggested in SM.

In deriving Eq.~(\ref{f-fact}) we assumed $J_{\bot}=$ 
const. Recognizing
now that   $J_{\bot}$ will be time-dependent, we 
note that disorientation of the grain depends primarily on the value of 
$J_{\bot}^{-1}$ near the time of crossover. We 
therefore seek to establish $\langle J_{\bot}^2(0)\rangle^{1/2}$, 
the value at the moment ($t=0$) of crossover.
One can write
\beq
\frac{{\rm d}\langle J_{\bot}^{2}-J_{\bot 0}^2\rangle}{{\rm d}t}=
N_{1}\langle (\bigtriangleup J_{\bot})^{2}\rangle
-\frac{2\langle J_{\bot}^{2}-J_{\bot 0}^2\rangle}{t_{B}}~~~,
\label{m22}
\eeq
which generalizes eq.~37 in SM.
All the time during a crossover, apart from a short interval when the grain
actually flips over, $\omega\gg \omega_{\bot}$ and 
 therefore it is possible to assume that 
$\omega_{z}\approx\omega$ (see SM).
According to our initial assumption, 
regular 
torques dominate the zero-order dynamics. Thus ${\rm d}t= 
(I_{z}/L_{z})\times {\rm d}\omega_{z}= (I_{z}/N_{1}\langle 
\bigtriangleup J_{z}\rangle){\rm d}\omega_{z}$ follows from 
Eq.~(\ref{m19}). Substituting
\beq
\zeta=\frac{\langle \bigtriangleup J_{z}\rangle}
	 {I_{z}[\langle (\bigtriangleup J_{\bot})^{2}_{\H}\rangle+
          N_{2}/N_{1}\langle (\bigtriangleup J_{\bot})^{2}_{g}\rangle]}
\left(\frac{2I_{z}}{A_{a}N_{1}|\langle \bigtriangleup J_{z}\rangle|}\right)^{1/3}
\langle J_{\bot}^{2}-J_{\bot 0}^2\rangle
\eeq
and
\beq
u\equiv \left(\frac{2I_{z}}{A_{a}N_{1}
\langle \bigtriangleup J_{z}\rangle}\right)^{1/3}\omega_{z}
\eeq
into Eq.~(\ref{m22}) gives (SM)
\beq
\frac{{\rm d}\zeta}{{\rm d}u}=1-\zeta u^{2}.
\eeq
Therefore for negative $\omega_{z}$ increasing to zero for $t=0$, one gets
\beq
\zeta(0)=3^{1/3}\Gamma\left(\frac{4}{3}\right)
\eeq
and 
\beq
\langle J_{\bot}^2(0)\rangle =J^2_{\bot,\it torque}+J^2_{\bot 0}~~~,
\eeq
where 
\begin{eqnarray}
J_{\bot, \it torque}^{2}&=&\frac{3^{1/3}\Gamma(4/3)}{2^{1/3}}
\frac{\omega_{zc}I_{z}}{|\langle \bigtriangleup J_{z}\rangle|}
\left[\langle (\bigtriangleup J_{\bot})^{2}_{\H}\rangle+\frac{N_{2}}{N_{1}}
\langle (\bigtriangleup J_{\bot})^{2}_{g}\rangle\right]\nonumber\\
&\approx&\left(\frac{3}{2}\right)^{1/3}
\Gamma\left(\frac{4}{3}\right)(A_{a}N_1)^{1/3}
|\langle \bigtriangleup J_z \rangle |^{-2/3} I_z^{2/3}
\langle(\bigtriangleup J_{\bot})^{2}_{\rm H}\rangle(1+\chi) ~~~. 
\label{a4}
\end{eqnarray}
and 
\beq
\chi\equiv \frac{N_{2}}{N_{1}}
\frac{\langle (\bigtriangleup J_{\bot})^{2}_{g}\rangle}
     {\langle (\bigtriangleup J_{\bot})^{2}_{\H}\rangle}~~~,
\eeq
$\Gamma(x)$ is the gamma function, and
\beq
\omega_{zc}=\left(\frac{A_{a}N_{1}|\langle \bigtriangleup J_{z}\rangle|}
			   {I_{z}}\right)^{1/3}
\eeq
is the value of $\omega_{z}$ such that the crossover time
 $(I_{z}\omega_{z})/
(N_{1}\langle \bigtriangleup J_{z}\rangle)$ equals the relaxation time. 
During this time the number of torque events is $|\omega_{z}|I_{z}/
\langle \bigtriangleup J_{z}\rangle$, and the product of this 
number over the mean squared increment of angular momentum per torque
event, i.e.
 $[\langle (\bigtriangleup J_{\bot})^{2}_{\H}\rangle+N_{2}/N_{1}
\langle (\bigtriangleup J_{\bot})^{2}_{g}\rangle]$, provides the 
estimate for  $\langle J_{\bot}^{2}-J_{\bot 0}^2\rangle$ in 
accordance with Eq.~(\ref{a4}).

Comparison of Eq.~(\ref{a4}) and eq.~(42) in 
SM shows that the mean value of $J_{\bot}^2$ 
arising from
stochastic torques
is increased by a factor $(1+\chi)$, 
where 
\beq
\chi \approx \frac{2k(T+T_d)}{\gamma E} 
\left(\frac{1.2-0.293y}{1-y}\right)
\approx 0.43
\left(\frac{T+T_d}{100K}\right)\left(\frac{0.2 {\rm eV}}{E}\right)
\hat{\gamma}^{-1}
\left(\frac{1.2-0.293y}{1-y}\right)~~~,
\label{m32}
\eeq
and $T$ is the gas temperature. In molecular gas with
$1-y\ll1$, $\chi$ can be large, but in HI regions $\chi\le 0.5$.
We also note that
$\chi$ does not depend on grain geometry.

Supersonic
drift causes mechanical alignment that
we briefly discuss in section~5.2. Here we limit discussion to the case where
gaseous bombardment is isotropic during the crossover event.

 For typical interstellar conditions, the 
$J_{\bot 0}^2$ term  in Eq.~(\ref{a4})
is much more
important than the term due to gaseous bombardment.
The importance of the Barnett fluctuations relative to the
stochastic torques is measured by
\beq
R^2=\frac{J_{\bot 0}^2}{J_{\bot, torque}^2}\approx 412\frac
{\hat{T}_d^{2/3}}{\hat{\gamma}_1^{1/3}(1-y)^{1/3}\hat{E}^{2/3}\hat{T}^{1/6}
\hat{n_{\H}}^{1/3}\hat{a}^{5/3}\hat{\alpha}^{1/3}}\frac{1}{1+\chi}\hat{K}_3(r)
~~~,
\label{ratio}
\eeq
where
\beq
\hat{K}_3(r)=\frac{9^{1/3} \times 41 r^{2}}{(4r^2-1)^{2/3}[16r^2(r+1)+6r+3]}~~~.
\eeq

The fact that the latter function tends to infinity for cubic grains ($r=1/2$)
is the consequence of the simplifications within our
 model. In fact, for cubic grains
the perpendicular component of the grain angular moment will be
of the order of the overall angular momentum, as pointed out
in \S 2. It is easy to see,
that for moderately oblate grains, however, $\langle J^2_{\bot}(0)\rangle$ is 
dominated by the
term $ J^2_{\bot 0}$ arising from
 thermal fluctuations. As $\chi\sim(1-y)^{-1}$, for 
small concentration of atomic hydrogen $J^2_{\bot, torque}$ may become
important. One of the problems with considering very small concentrations
of molecular hydrogen is that disorientation then happens not only during
crossovers but also during spin-ups (see Lazarian 1995d) 
and this requires the theory of 
suprathermal alignment to be modified. Moreover, according to Eq.~(\ref{a})
for $y\rightarrow 1$ only very large grains obey the theory we discuss
here.

On estimating the critical size of $a_c$ in Eq.~(\ref{a}) we assumed
that $\langle J_{\bot}^2\rangle=J^2_{\bot 0}$. In general, this is
not true and our estimate of $a_c$ should be multiplied by $(1+R^{-2})^{1/5}$.
The latter value, however, is $\sim 1$ according to Eq.~(\ref{ratio})
and therefore our estimate  of the critical size given by Eq.~(\ref{a}) 
stays essentially unaltered.

Although in the grain frame of reference $J_{\bot, torque}$  and
$J_{\bot 0}$ appear very similar, their difference is obvious in
the inertial frame. Indeed, $J_{\bot 0}$ that arises from thermal
fluctuations within the grain material does not alter the direction 
of $\bf J$ in the latter frame. On the contrary, $J_{\bot, torque}$
that arises from gaseous bombardment and stochastic events of
H$_2$ formation does directly affect the direction of $\bf J$.

As seen from Eq.~(\ref{ratio}), we expect to have $J_{\bot 0}^2 \gg 
J_{\bot,torque}^2$; in this limit,  the disorientation
 parameter $F$ (see Eq.~(\ref{f-fact})) can be obtained:
\beq
F\approx 9.0\times 10^{-3} \hat{E}^{1/2}\hat{\alpha}^{1/2}\hat{a}^{-1/2}
\hat{T}_d^{-1/2}
\hat{\rho_s}^{-1/2}(1+\chi)\hat{K}_4(r) ~~~,
\label{a101}
\eeq
where
\beq
\hat{K}_4(r)=\sqrt{\frac{5}{3}}
\frac{8(2r+5)(4r^2-1)^{1/2}}{179 r^{1/2}(4r^2+1)^{1/2}}\left(1+\frac{3(16r^2(r+1)+6r+3)}{8r^2(2r+5)}\right)~~~.
\eeq
The disorientation decreases
 as $r\rightarrow 1/2$, which corresponds to a cubic grain.

For the case of $J_{\bot,torque}^2 \gg J_{\bot 0}^2$
we can also obtain an  estimate of $F$:
\beq
F\approx 0.26~\hat{\alpha}^{1/3}
\hat{\rho_s}^{-1/2}\hat{a}^{-4/3}
(\hat{n}_{\H}\hat{E}\hat{\gamma}_{1}\hat{T_d})^{-1/6}\hat{T}^{1/12}
\sqrt{1+\chi}
\hat{K}_5(r)~~~.
\label{a12}
\eeq
where 
\beq
\hat{K}_5(r)=\frac{8}{179}\frac{\sqrt{205}(2r+5)(4r^2-1)^{1/6}r^{1/2}}
{ 3^{1/6}(1+4r^2)^{1/2}(16r^2(r+1)
+6r+3)^{1/2}}\left(1+\frac{3(16r^2(r+1)+6r+3)}{8r^2(2r+5)}\right)~~~.
\eeq
 This estimate
coincides with that in 
Lazarian (1995c) in the limit of negligible contribution from the
gaseous bombardment\footnote{The difference in the numerical
values obtained here and in Lazarian (1995c) stems from the fact that 
in the latter paper the function $F$ was defined as the average value
for an ensemble of grains with varying $J_{\bot}$ and $\nu$ (see \S 4.1).
}.

\section{Implications for the alignment}
\subsection{Paramagnetic alignment}

Paramagnetic alignment of suprathermally rotating grains -- frequently
called  Purcell alignment -- can be described using the equation 
(Purcell 1979):
\beq
\frac{{\rm d}\theta}{{\rm d}t}=-\frac{\sin\theta \cos\theta}{t_r}~~~,
\label{58}
\eeq
where $t_r$ is the time of relaxation time of a grain with volume $V$
in the ambient field $B$:
\beq
t_r=\frac{I_z}{B^2VK}=6.7 \times 10^{13}
\frac{\hat{\varrho}_s\hat{a}^2 r^2 \hat{T}_d}{\hat{B}^2} ~~~{\rm s},
\eeq
where $K\approx 1.2 \times 10^{-13}\hat{T}_d^{-1}$~s (Draine 1996).

The solution of the differential equation above is trivial:
\beq
\tan\theta=\tan\theta_0 \exp(-t/t_r)~~~.
\label{triv}
\eeq

If at $t=0$ grains are initially randomly oriented, then after time $t$
we obtain the Purcell (1979) expression for 
\beq
Q(t)\equiv 3/2([\cos^2\theta]-1/3)=\frac{3}{2}\frac{1-(e^{\delta}-1)^{-1/2}\arctan\sqrt{e^{\delta}-1}}
{1-e^{-\delta}}-\frac{1}{2}~~~,
\eeq
where $\delta\equiv 2t/t_r$ and square brackets denote averaging
over grain initial orientations.

Now suppose that grains are randomly oriented following crossovers
and let $P(t)dt$ be the probability that a randomly selected
grain will have gone a time $t_b\in[t, t+dt]$ since its last 
crossover event.
To obtain the Rayleigh reduction factor (Greenberg 1968)
\beq
\sigma=\frac{3}{2}(\langle\cos^2\theta\rangle -1/3)~~~,
\label{r-fact}
\eeq
one needs to average $Q(t)$:
\beq
\sigma=\frac{\int_0^\infty 
              Q(t)P(t)d t}
            {\int_0^\infty 
              P(t)d t}~.
\label{62}
\eeq

For our simplified treatment we will assume that for 
{\it any particular grain} 
in the ensemble the crossovers happen periodically with period
$t_{\rm max}$. \footnote{The theory of 
alignment for an arbitrary distribution of time 
intervals between crossovers is given in Draine \& Lazarian (1996).}
Then 
\beq
P(t_{\rm max})=\left\{
\begin{array}{l}
t^{-1}_{\rm max},~~~t<t_{\rm max}~,\\
0,~~~t>t_{\rm max}~,
\label{pmax}
\end{array}
\right. 
\eeq
For this distribution the mean time between crossovers (or zero-crossings) is 
$\bar{t}_z=t_{\rm max}$.
Integrating (\ref{62}) we get
\beq
\sigma= 1+\frac{3}{\delta_{\rm max}}\left[
\frac{{\rm arctan}\sqrt{e^{\delta_{\rm max}}-1}}
     {\sqrt{e^{\delta_{\rm max}}-1}}-1\right]~~~,
\label{sig}
\eeq
where $\delta_{\rm max}=2\bar{t_z}/t_r$.
For small $\delta_{\rm max}$ Eq.~(\ref{sig}) can be expanded 
\beq
\sigma \approx \frac{\delta_{\rm max}}{10}+\frac{\delta_{\rm max}^2}{210}
+\frac{\delta_{\rm max}^3}{840}+\dots ~.
\eeq
Up to now we assumed complete disorientation in the course of a crossover.
 It is 
evident from 
Table~2
that $t_r\gg t_d$ for typical interstellar conditions.
As the mean ``time back to crossover'' for ``short-lived spin-up''
(e.g. for $t_L<t_d$) is of the order of $t_d$, paramagnetic alignment is
marginal unless  the directions of $\bf J$ before and after crossovers are
strongly correlated (SM).

To account for incomplete disorientation SM adopted the following
reasoning: consider crossovers that occur at intervals $t_{\rm max}$; then
in a time $t_{\rm max}/F$ the disorientation decreases $\cos\eta$
by $1/e$. Thus,  
according to SM, the effects of incomplete disorientation during crossovers
may be approximated by replacing $\bar{t}_z$ by 
$\bar{t}_z/{\rm min}[\langle F\rangle_J, 1]$. Henceforth 
we use 
 $\langle ..\rangle_J$ to denote averaging over the distribution
of $J_{\bot}$. We remind our reader that up to now we evaluated $F$
for a grain with $J_{\bot}=\langle J_{\bot}^2 \rangle^{1/2}$.

We conjecture that  replacing $\delta_{\rm max}$ in Eq.~(\ref{sig})
\beq
\delta_{\rm eff}=\frac{2\bar{t}_z/t_r}{(1-\exp(-\langle F \rangle_J))}
\label{x-fact}
\eeq
to obtain
\beq
\sigma \approx 1+\frac{3}{\delta_{\rm eff}}\left[
\frac{{\rm arctan}\sqrt{e^{\delta_{\rm eff}}-1}}
     {\sqrt{e^{\delta_{\rm eff}}-1}}-1\right]~~~,
\label{sig1}
\eeq
 may give a
better fit than the SM approximation above, as it allows for 
residual correlation for $\langle F\rangle_J \gtrsim 1$. It is evident that 
for $\langle F\rangle_J \gg 1$ and $\langle F\rangle_J \ll 1$ 
our approximation coincides with that in SM. 

To study the effects of incomplete disorientation below we use Monte-Carlo
simulations to calculate $\sigma$ for different ratios of $\bar{t}_z/t_r$ and
 $\langle F\rangle_J$.

To obtain $\langle F\rangle_J$, we require 
$\langle 1/J_\bot(0)\rangle_J$.
To estimate this we note 
that $\langle J_{\bot}^{2}(0)\rangle=\langle J_{x}^{2}(0)\rangle+\langle J_{y}^{2}(0)\rangle$ and due to 
the symmetry inherent to the problem
\beq
\langle J_{x}^{2}(0)\rangle=\langle J_{y}^{2}(0)\rangle
=0.5\langle J_{\bot}^{2}(0)\rangle~~~.
\label{a7}
\eeq
For a Gaussian distribution with $\sigma_{1}^{2}=\langle J_{x}^{2}(0)\rangle$, 
one gets
\beq
\left \langle\frac{1}{ J_{\bot}(0)}\right \rangle_J
= \frac{1}{2\pi \sigma_{1}^2}
\int_{0}^{+\infty}2\pi r \frac{1}{r}
\exp\left\{-\frac{r^2}{2\sigma_{1}^{2}}\right\}
{\rm d}r=\sqrt{\frac{\pi}{2}}
\frac{1}{\sigma_{1}}=\frac{\sqrt{\pi}}{\langle J_{\bot}^{2}(0)\rangle^{1/2}}~~~.
\label{a8}
\eeq

Thus from Eq.~(\ref{f-fact}) we get
\begin{eqnarray}
\langle F\rangle_J &\approx & \frac{\pi^{3/2}}{4}K_1(r)
\frac{\langle(\bigtriangleup J_{z})^{2}\rangle}
     {|\langle\bigtriangleup J_{z}
      \rangle|\langle J_\bot^2(0)\rangle^{1/2}}\nonumber\\
&=&
1.60 
\times 10^{-2} \hat{E}^{1/2}\hat{\alpha}^{1/2}\hat{a}^{-1/2}
\hat{T}_d^{-1/2}
\hat{\rho_s}^{-1/2}(1+\chi)\hat{K}_5(r)(1+R^{-2})^{-1/2}.
\label{a10}
\end{eqnarray}

In those simulations we use Eq.~(\ref{m17}) to find 
 $\langle\cos\eta_f\rangle$ for a fixed $J_{\bot}$ and then
perform numerical averaging of  $\langle\cos\eta_f\rangle$ over
a Gaussian  distribution of $J_{\bot}$. We require 
a distribution function for  the
stochastic jumps that correspond to  
$\langle\langle\cos\eta_f\rangle\rangle_{J}$.
We assume the distribution of $\eta$ to have the form\footnote{
The assumed functional form (\ref{p-distr}) has the required behavior
of $dP_1/d\eta=0$ for $\eta=0, \pi$; $P_1\sim (\sin\eta)$ for
$\alpha \rightarrow 0$; $P_1 \sim \eta \exp(-\alpha^2 \eta^2/4)$  for 
$\eta \ll 1$.}
\beq
P_1(\eta){\rm d}\eta=C\sin\eta\exp(-\alpha^2\sin^2(\eta/2)){\rm d}\eta~~~,
\label{p-distr}
\eeq
where $C$ is the normalization constant:
\beq
C=\frac{\alpha^2}{2} \frac{1}{1-\exp(-\alpha^2)}
\eeq
and $\alpha$ is the solution of the transcendental equation:
\beq
\langle\cos\eta_f\rangle=\frac{1+\exp(-\alpha^2)}
{1-\exp(-\alpha^2)}-\frac{2}{\alpha^2} ~~~.  
\eeq

An individual jump over $\eta$ during a crossover event happens in
a random direction and we obtain the final value of $\theta_{i,f}$ after the
$i$-th crossover from the following formulae:
\beq
\cos\theta_{i,f}=\cos\theta_{i,b}\cos\eta+\sin\theta_{i,b}\sin\eta\cos x ~~~,
\eeq
where $x$ is a random variable uniformly distributed over 
$[0,2\pi]$ and $\theta_{i, b}$ is the value of the alignment angle
just before the $i$-th crossover. Between crossovers
the dynamics of the alignment angle 
is determined by Eq.~(\ref{triv}). Averaging over $P(t_{\rm max})$
(see Eq.~(\ref{pmax})) is also performed to account for the distributions of
the times since the last crossover.

The results of those calculations 
are shown in   Fig.~2 where we have plotted $\sigma$ vs $\delta_{\rm eff}$
(defined by Eq.~(\ref{x-fact})).  
For each value of $\delta_{\rm eff}$ different symbols correspond
to the alignment measures obtained for different values of 
$\langle F \rangle_J$. The solid
line in the same plot corresponds to Eq.~(\ref{sig1}).
In the limit $\langle F \rangle_J \rightarrow \infty$ we have complete 
disorientation, in which case Eq.~(\ref{sig1})
is an exact result (for periodic crossovers). However, it is evident 
that Eq.~(\ref{sig1})
provides a good approximation to the numerical results for finite 
$\langle F \rangle_J$, at least for periodic crossovers.
More general models where the times between crossovers are
obtained through Monte-Carlo simulations
are studied in Draine \& Lazarian (1997). 

For typical values of interstellar parameters (see Table~2) one obtains
$F\approx 0.014$ (see Eq.~\ref{a101}). Using our earlier estimate
$t_z\approx 1.6 t_d\hat{a}^{1/2}$ (for assumed $t_L/t_d=0.25/\hat{a}$)
we get $\delta_{\rm eff} \approx 13.7$  (indicated by an arrow in Fig.~2), 
for which Eq.~(\ref{sig1}) gives
$\sigma\approx 0.8$. This high degree of
alignment is due to the small estimated value of $\langle F \rangle$.
In fact, it was argued
in Lazarian (1995c) that $t_L$ is expected to be several times
greater than $t_d$ for grains with $a>10^{-5}$ cm and this, by
increasing $t_z$,
would  further increase  the expected alignment.

Although we do not quantitatively
discuss here the alignment of grains with $a<a_c\approx 1.5\times 10^{-5}$~cm 
we conjecture that the alignment of such grains may be suppressed by the
effective disalignment during each crossover (i.e. $\langle F \rangle \gg 1$),
 due to
the variations in the angle $\beta$ due to Barnett fluctuations when
$t_B\ll t_c$ (see \S 3.2). The strong dependence of $t_B/t_c$ on $a$ 
(see Eq.(\ref{tbtc}))
suggests that this may account for the observed lack of alignment of
interstellar grains with $a\lesssim 10^{-5}$ (Kim \& Martin 1995).

We see, then, that if the only important torques were
those due to H$_2$ formation, gas-grain collisions, and 
paramagnetic dissipation, we would expect paramagnetic grains in diffuse
clouds to be substantially aligned for $a>a_c\approx 1.5\times 10^{-5}$ cm,
and probably minimally aligned for $a<a_c$, at least qualitatively 
consistent with observations. It has recently been recognized, however,
that  starlight plays a major role in the dynamics of
$a\gtrsim 0.1$ $\mu$m grains: the torques exerted by anisotropic starlight
($i$) drive suprathermal rotation (Draine \& Weingartner 1996) and
($ii$) can directly act to align $\bf J$ with the interstellar magnetic field
 (Draine \& Weingartner 1997).

Since we have neglected starlight torques in the present paper, our
conclusion for $a\gtrsim 0.1$ $\mu$m grains is only preliminary.
A study of crossovers incorporating the effects of both H$_2$ formation
and anisotropic starlight is planned. It appears
to us highly likely that when both effects are included, the observed
lack of alignment of $a\lesssim 0.1$ $\mu$m grains, and substantial alignment 
for  $a\gtrsim 0.1$ $\mu$m grains, will be explained. 

We recall that suprathermal rotation can also be driven by variations
of the accommodation coefficient and photoelectric emissivity (Purcell 1979).
In a molecular ($y=1$) region with no ultraviolet light, Purcell's
estimate for the torque due to variations in accommodation
coefficient leads to $a_c=3.0\times 10^{-5}$ cm as the radius for 
which $t_B=t_c$. However, for this case we also find $\langle F \rangle >1$
for $a\approx a_c$, so that Purcell alignment will be insufficient 
unless $\bar{t}_z\geq t_r$.

\subsection{Mechanical alignment}

It was previously thought
that suprathermally rotating grains are not subject to mechanical
 alignment when the gas is streaming relative to the grain. 
However, two mechanisms of mechanical alignment of suprathermally
rotating grains, namely, ``cross-section'' and ``crossover'' alignment,
were proposed by Lazarian (1995d). The first process is caused by the
fact that the frequency of crossover events depends on the value of
the cross-section exposed to the gaseous flux (see Lazarian \& Efroimsky
1996 for more details). The second mechanism arises from the substantial
susceptibility of grains to alignment by gas-grain streaming during 
crossover events. 

Both mechanical processes
are related to the phenomenon of crossovers. Thus our finding of
reduced disorientation during crossovers is a new feature that should
be incorporated into the discussion of mechanical alignment.
As we mentioned earlier, this reduced randomization is valid only for
 grains with $a>a_c$, where $a_c$ is given by Eq.~(\ref{a}),
 and therefore no changes of the earlier results are 
expected for grains with $a<10^{-5}$~cm. Such grains can be aligned,
for instance, by ambipolar diffusion, which favors small
grains.

To start with, consider the cross-section mechanism. In 
Lazarian (1995d)
 this mechanism was exemplified using a toy model, namely,
a flat disc grain which randomly jumps in the course of a crossover from 
one position,
where the surface of the disc is parallel to the flow, to the other
position, where the disc surface is perpendicular to the flow. If the
probability per unit time of a crossover is proportional to
 the rate at which atoms arrive at the
grain surface, it is easy to see  that the grain will spend more time at
 orientations where the cross section presented to the streaming gas
is minimal. Within the toy model above,
this corresponds to the position with 
the surface of the disc parallel to the flow.

In other words, the cross-section mechanism uses the fact that $\bar{t}_z$ is
a function of the angle $\phi$ between the grain axis of major inertia
and
the direction of the gaseous flow. Roughly speaking, our study above
shows that crossovers with disorientation parameter $\langle F \rangle$
and mean time between crossovers $\bar{t}_z$ are equivalent to
crossovers with complete disorientation and the mean time between crossovers
$\bar{t}_z/(1-\exp(-\langle F \rangle))$. If 
$F$ is dominated by thermal fluctuations its dependence on
gaseous bombardment vanishes (see Eq.~(\ref{a101})) as does its dependence
on $\phi$. Thus the only effect of incomplete disorientation during
crossovers (as compared to full disorientation) is to increase
of alignment time  (the time to attain a steady-state) 
by a factor $(1-\exp(-\langle F \rangle))^{-1}$.

``Crossover alignment'' depends on the ratio of the randomizing torques
arising from H$_2$ formation and aligning torques caused by gaseous
bombardment (see Lazarian 1995d). This ratio neither depends on
the number of crossovers nor on the time of alignment. The fact that
the thermal fluctuations do not change the direction of $\bf J$ is
essential for understanding why this type of alignment is not
suppressed in the presence of the incomplete disorientation during
crossovers. It is possible
to show, however, that the time of alignment increases by a factor
 $(1-\exp(-\langle F \rangle))^{-1}$.

In spite of the fact that the measure of the mechanical alignment
does not change, our observation 
 that the time required to reach steady state is increased by a
factor $(1-\exp(-\langle F \rangle))^{-1}$ can be important.
This is particularly important whenever
grain alignment is caused by a transient phenomenon, e.g., a 
MHD shock. If $\bar{t}_z/(1-\exp(-\langle F \rangle))$ is much 
longer than the time of streaming,
the alignment of grains with $a>1.5\times 10^{-5}$ cm will be  
marginal\footnote{This by no means preclude small grains from being aligned
and we believe that the preferential alignment of small grains could be
a signature of such an alignment.}.

\section{Conclusion}

We have shown that thermal fluctuations within the grain 
material limit the extent to which
 the axis ${\bf a}_1$ 
of major inertia can be aligned with the angular momentum $\bf J$ 
in suprathermally rotating grains.
Although the fluctuating angle $\beta$ between  ${\bf a}_1$ and  $\bf J$ 
is tiny
when the grain is rotating suprathermally, it becomes larger and of 
critical importance
during periods of crossovers.
We have proved that for  grains with $a>a_c\approx 1.5\times 10^{-5}$ cm
the non-zero component of ${\bf a}_1$  
 perpendicular to  $\bf J$
arising from thermal fluctuations substantially {\it diminishes}
the degree of the randomization of the angular momentum direction
in the course of crossovers. If the only torques acting on a grain are
those due to gas-grain collisions, H$_2$ formation, and paramagnetic
dissipation, our estimates show that for large ($a\gtrsim a_c$) grains the 
grain alignment is close to perfect, while
small ($a<a_c$) would have only marginal alignment.

If there is gas-grain streaming, the thermal fluctuations
increase the time for mechanical alignment for large suprathermally
rotating grains, 
but do not alter the limiting
 steady state measure of alignment. If the mechanical alignment
is caused by Alfvenic waves, it acts in unison with the paramagnetic 
mechanism to enhance the alignment of large grains. For small grains 
mechanical alignment due to transient phenomena (e.g. ambipolar diffusion
within MHD shocks) can be the dominant cause of alignment. 

\acknowledgements
A.L. is extremely grateful to Lyman Spitzer for elucidating discussions of
the crossover effect.
A.L. acknowledges the support of NASA grant NAG5 2858 and B.T.D. the support
of NSF grant AST-9219283.

\appendix
\section{Some Results for a Square Prism}

Here we consider a square prism, 
with dimensions $b\times b\times a$, density $\rho_s$,
mass $\rho_s b^2 a$, area $4ba+2b^2$, and moments of inertia
$I_z=(8/3)\rho_s a (b/2)^4$ and $I_x=I_y=(1/3)\rho_sa(b/2)^2[a^2+4(b/2)^2]$.
We let $r\equiv (b/2a)$ and note that $r=1$ corresponds
to the prism grain discussed in Purcell (1979).
We consider a hydrogen-helium gas with density
$n_\H\equiv n(\H) + 2n(\HH)$, temperature $T$, 
molecular fraction
\beq
y \equiv {2n(\HH)\over n_\H}
~~~;
\label{ap1}
\eeq
and $n(\He)/n_\H=0.1$.
The square prism has
\beq
t_{M}={\rho_s a \over n_\H v_\H m_\H (1.2-0.293y)}\frac{2r}{(r+1)}
~~~,~~~
t_d = {(r+1)\over(r+2)}t_{M}
~~~,
\label{ap2}
\eeq
where
$t_{M}$ is the time for the grain to collide with its own mass of gas,
$t_d$ is the
rotational damping time
(assuming incident atoms to temporarily
stick),\footnote{
	Purcell \& Spitzer (1971) give $t_d/t_M$ for a square prism but
	their eq.(9)
	contains a typographical error: the factor $(5s+1)$ should instead
	be $(4s+1)$. Our $r$ is equal to $1/(2s)$ as defined in Purcell
        \& Spitzer (1971).
	}
and
$v_\H=(8kT/\pi m_\H)^{1/2}$ is the mean speed of H atoms.
If a fraction $\gamma$ of impinging H atoms are converted to H$_2$, then
the H$_2$ formation rate is
\beq
N_1 = r(r+1)\gamma (1-y)n_\H v_\H a^2 
~~~.
\label{ap3}
\eeq
We assume the grain to be spinning around the $z$-axis.
The prism is assumed to have $\nu$ active sites of H$_2$ formation distributed
randomly over the surface.
Following Purcell, we assume that
newly-formed H$_2$ molecules depart
from each recombination site at a rate $N_1/\nu$,
with fixed kinetic energy $E$ but
random directions ($dP/d\theta=2\sin\theta\cos\theta$, where $\theta$ is
with respect to the local surface normal).
The $\nu/(1+r)$ sites on the sides of the prism then 
produce a steady torque $L_z$ with
\beq
\langle L_z^2\rangle ^{1/2} =
r^2(r+1)^{1/2}
\gamma (1-y)n_\H v_\H a^3 \left({2m_\H E\over 3 \nu}\right)^{1/2}
~~~,
\label{ap4}
\eeq
and a mean angular impulse per recombination event $\langle\Delta J_z\rangle$,
with
\beq
|\langle\Delta J_z\rangle |\approx {\langle L_z^2\rangle^{1/2}\over N_1} = 
\frac{r}{(r+1)^{1/2}}\left( {2 m_\H a^2 E \over 3\nu} \right)^{1/2}
~~~.
\label{ap5}
\eeq
Individual H recombination events, occurring at a rate $N_1$, contribute
random angular momentum impulses with
\beq
\langle (\Delta J_z)^2\rangle_\H = \frac{1}{3}\frac{r^2(2r+5)}{r+1}m_\H a^2 E
~~~,
\label{ap6}
\eeq
\beq
\langle (\Delta J_\perp)^2\rangle_\H = \frac{16r^2(r+1)+6r+3}{12(r+1)}
m_\H a^2 E
~~~.
\label{ap7}
\eeq

Gas particles impinge at a rate 
\beq
N_2=2r(r+1)n_\H v_\H a^2(1.05-y+y/\surd8)
~~~,
\label{ap8}
\eeq
If impinging particles
temporarily stick and then thermally desorb at
temperature $T_d$, then these collision events produce
\beq
N_2\langle (\Delta J_z)^2\rangle_g = 
{2\over 3}r^3(2r+5) n_\H m_\H v_\H xa^4 k(T+T_d)
\left(1.2-y+{y\over\surd2}\right)
~~~,
\label{ap9}
\eeq
\beq   
N_2\langle (\Delta J_\perp)^2\rangle_g = 
{1\over6} r(16(r^2(r+1)+6r+3)n_\H m_\H v_\H a^4 k(T+T_d)
\left(1.2-y+{y\over\surd2}\right)
~~~.
\label{ap10}
\eeq

\pagebreak

\begin{table}
\begin{center}
\begin{tabular}{|c|c|}
\hline
notation                      & meaning\\
\hline
$y\equiv 2n(H_2)/n_H$         & H$_2$ fraction\\
$\gamma\equiv 0.2\hat{\gamma}$  & H recombination efficiency\\
$v_H\equiv (8kT_{gas}/\pi m_H)^{1/2}= 1.5\times 10^{5} \hat{v_H}$ cm s$^{-1}$&
thermal velocity\\
$a\equiv 10^{-5}$ cm & grain size\\
$r\equiv b/2a$       & grain axis ratio divided by 2\\
$E\equiv 0.2 \hat{E}$ eV & kinetic energy of nascent H$_2$\\
$\alpha\equiv 10^{11}\hat{\alpha}$ cm$^{-2}$ & surface density of recombination
sites\\
$\nu\equiv 80r(r+1)\hat{\alpha}\hat{a}^2$ & number of recombination sites\\
$T_d\equiv 15 \hat{T_d}$ K & grain temperature\\
$T\equiv 85 \hat{T}$ K     & gas temperature\\
$n_H\equiv 20 \hat{n}_H$  cm$^{-3}$ & density of H nucleon \\
$\varrho_s\equiv 3 \hat{\varrho}_s$ g cm$^{-3}$ & solid density\\
$B\equiv 5 \hat{B}\times 10^{-6}$ G & magnetic field\\
\hline
\end{tabular}
\end{center}
\caption{Parameters of grains and ambient medium adopted
in this paper.}
\end{table}

\begin{table}
\begin{center}
\begin{tabular}{|c|c|}
\hline
Crossover      & $t_c=2.9\times10^9(\hat{\rho}\hat{T}_d \hat{\nu}\hat{E}^{-1}
\hat{a}^{-1})^{1/2}/(\hat{\gamma}\hat{n}_H \hat{v}_H (1-y))\hat{Z}_0(r)$ s\\
Barnett effect & $t_B=7.1\times 10^{7}\hat{a}^7\hat{\rho}_s^2
(\omega_{\bot 0}/\omega)^2 \hat{Z}_{1}(r)$ s\\
Gaseous damping &$t_d=3.3\times 10^{12}\hat{a}\hat{\rho_s}
(\hat{n_H}\hat{v_H})^{-1}(1.2-y+y/\sqrt{2})\hat{Z}_2(r)$ s\\
Paramagnetic relaxation &$t_r=6.7\times 10^{13}\hat{\rho_s}\hat{a}^2\hat{T_d}
\hat{B}^{-2}\hat{Z}_3(r)$ s\\
\hline
\end{tabular}
\end{center}
\caption{Characteristic times involved. We use $K\approx 1.2\times 
10^{-13}\hat{T}_d^{-1}$ for paramagnetic relaxation
(Draine 1996). The functions of
grain axis ratio $\hat{Z}_i$
$i=0,3$ are as follows: 
$\hat{Z}_0(r)=(6(1+4r^2))^{1/2}/(5(4r^2-1)(r+1))^{1/2}$, 
$Z_1(r)=(4r^2+1)^2/(25r^2)$, $Z_2(r)=2r/(r+2)$ and $Z_3(r)=r^2$.}
\end{table}

\clearpage

\begin{figure}
\begin{picture}(441,216)
\includegraphics{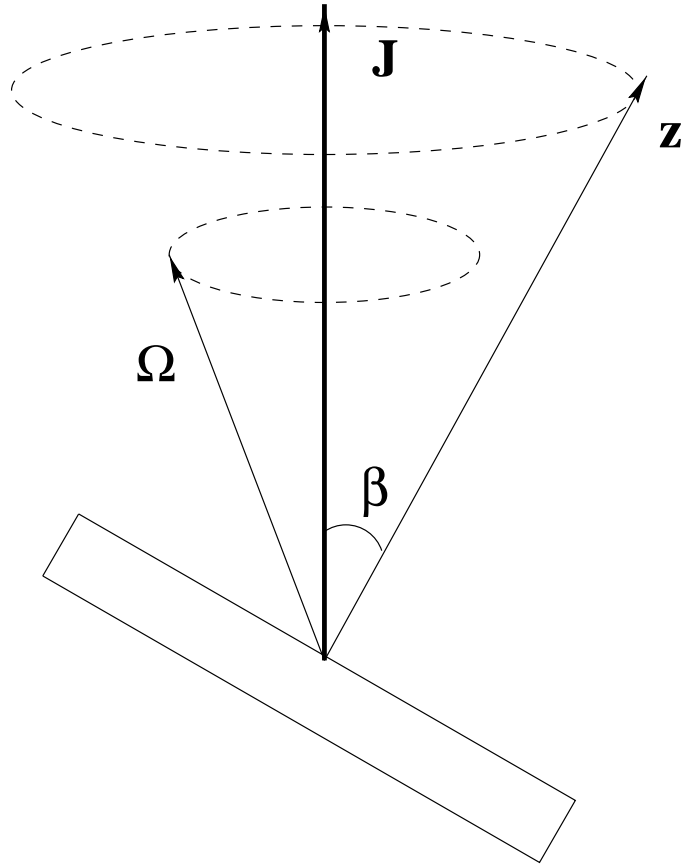}
\end{picture}
\caption[]{Grain's axis of major inertia $\bf z$ and $\bf \Omega$ 
precess about the direction of angular momentum $\bf J$.}
\end{figure}

\clearpage

\begin{figure}
\begin{picture}(441,216)
\includegraphics{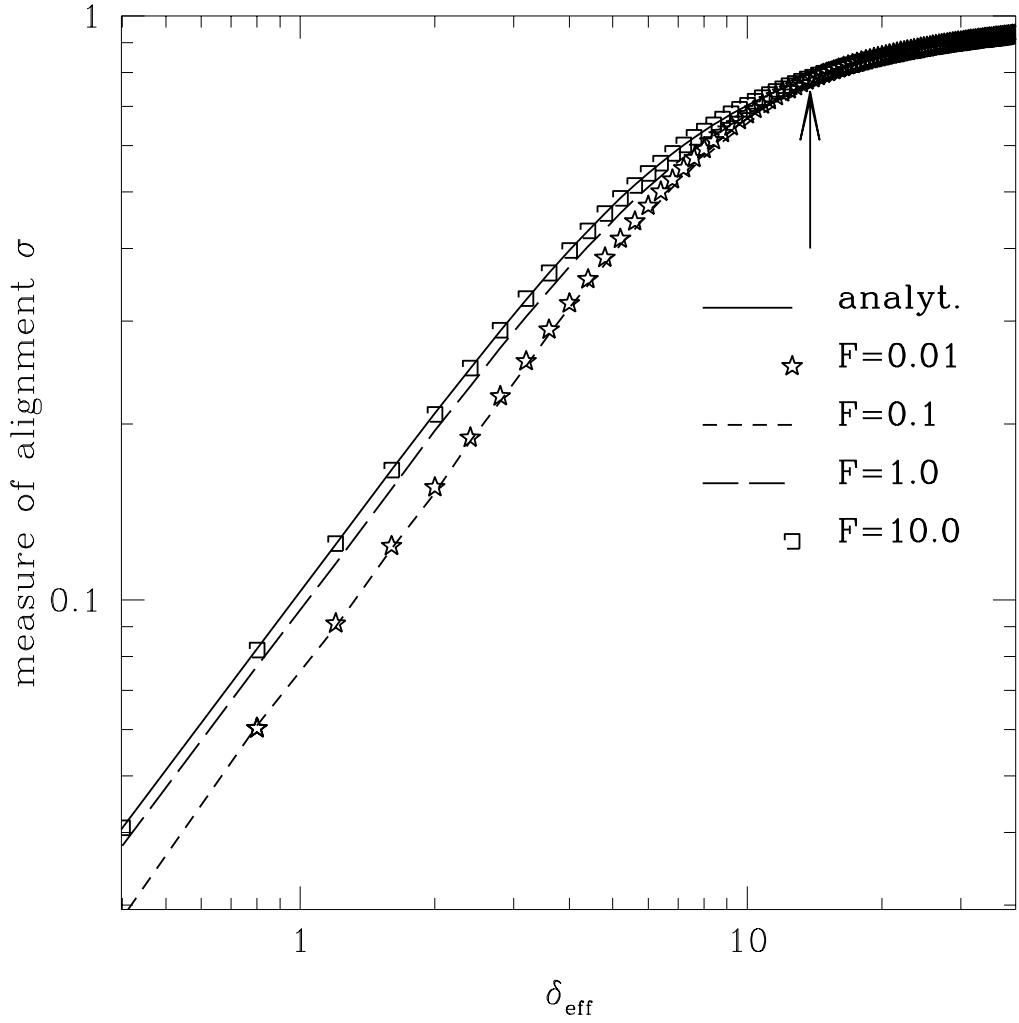}
\end{picture}
\caption[]{Measure of alignment $\sigma$ [see Eq.~(\ref{r-fact})] for 
periodic crossovers as a 
function of $\delta_{\rm eff}$, defined by Eq.~(\ref{x-fact}). 
The solid line is the analytic estimate (\ref{sig1}).The arrow corresponds to 
$\delta_{\rm eff}$ estimated for typical interstellar conditions (see text).}
\end{figure}

\end{document}